\documentclass[twocolumn,aps,pre,groupedaddress,amsmath,amssymb]{revtex4}

\input{epsf.tex}

\begin{document}


\title{Autowaves in the model of avascular tumour growth}


\author{A.V. Kolobov}
\email[]{kolobov@lpi.ru}
\affiliation{P.N. Lebedev Physical Institute of Russian Academy of
Sciences, 53 Leninsky prospect Moscow, 119991 RUSSIA}
\author{V.V. Gubernov}
\email[]{gubernov@lpi.ru}
\affiliation{P.N. Lebedev Physical Institute of Russian Academy of
Sciences, 53 Leninsky prospect Moscow, 119991 RUSSIA}
\author{A.A. Polezhaev}
\email[]{apol@lpi.ru}
\affiliation{P.N. Lebedev Physical Institute of Russian Academy of
Sciences, 53 Leninsky prospect Moscow, 119991 RUSSIA}


\date{\today}

\begin{abstract}
A mathematical model of infiltrative tumour growth taking into
account cell proliferation, death and motility is considered. The
model is formulated in terms of local cell density and nutrient
(oxygen) concentration. In the model the rate of cell death
depends on the local nutrient level. Thus heterogeneous nutrient
distribution in tissue affects tumour structure and development.
The existence of automodel solutions is demonstrated and their
properties are investigated. The results are compared to the
properties of the Kolmogorov-Petrovskii-Piskunov and Fisher
equations. Influence of the nutrient distribution on the autowave
speed selection as well as on the relaxation to automodel solution
is demonstrated. The model adequately describes the data, observed
in experiments.
\end{abstract}


\maketitle

\section{Introduction} \label{intro}

According to experimental data tumour growth can be naturally
subdivided into two stages \cite{Araujo04}. The initial stage,
called avascular growth, is characterized by the tumour
consumption of crucial nutrients, such as oxygen or glucose,
basically via diffusion. Depending on nutrients concentration
tumour cells are supposed to be in one of the three states:
proliferating, resting or dead. An avascular tumour is a compact
spherically symmetric colony of cancer cells with the necrotic
region in the center. Such a rigid structure holds until the
tumour size does not exceed several millimeters. Later in response
to chemical signals from cancer cells new capillary blood vessels
start to grow to provide better nutrients supply. This process,
called angiogenesis, indicates the second phase of tumour
development -- vascular growth.

Although a significant progress in experimental methods has been
achieved by now, they still cannot give a conceptual framework
within which all the existing data of the tumour development can
be fitted \cite{Gatenby_03}. Thereupon the interest in
mathematical modelling of the tumour growth and progression has
rapidly grown in the last decades. The majority of these models
consider avascular tumour growth or growth of multicellular tumour
spheroids (MTS) which are prevalent experimental {\em in vitro}
models of avascular tumours.

There are several basic classes of mathematical models for the
avascular tumour or MTS growth. One of them corresponds to
``reaction-diffusion type'' models which generally consist of an
ordinary differential equation for the tumour volume, coupled to
one or more parabolic partial-differential equations describing
the distribution of nutrients and growth inhibitory factors within
the tumour \cite{Greenspan_72, Byrne_96, Bueno_05}. In models of
this type nutrient concentration provides a mitotic and/or death
index and the equation for the tumour volume follows from the mass
conservation law. Thus, reaction-diffusion models do not consider
any cell motion. Also these models can deal only with spherically
or cylindrically symmetric spatial structure of a tumour.

In convection-domination models a tumour is considered as
incompressible fluid where cell motion is determined by convective
velocity field \cite{ Ward_97, Convection-1, Tao05}. In the models
of this type several convection equations describe dynamics of
different tumour cell types or phases. Local changes in cell
population lead to variations in the internal pressure, which in
turn induces motion of tumour cells. These models also include
reaction-diffusion equations for spatial nutrient and drug
distribution within the tumour. However convection-domination
models usually neglect tumour own cell motility. There are rather
few models which take into account both convective and random
tumour cell motion as well as chemo- or haptotaxis \cite{
Kolobov01, Kolobov03}.

For simulation of tumours with large random cell motility
Fisher-like equations have usually been used \cite{ Traqui_95,
Sherratt_01}. In these models a logistic shape for the tumour cell
proliferation rate is often used to prevent unlimited growth of
local cell density. This rather artificial approach does not take
into account nutrient distribution inside the tumour and completly
disregards the main cancer property -- unlimited growth in the
case of a sufficient level of nutrients.

With a rare exception the majority of papers on cancer simulation
consider solid tumours as compact objects with the total tumour
cell density close to the maximal possible cell density in the
tissue. However there is another tumour type, namely infiltrative
tumours, for instance glioma, characterized by a rather low value
of tumour cell relative density and a large volume of penetration
in normal tissue provided by high individual cell motility
\cite{Glioma_invasion}. There are several mathematical models
which describe tumours of this type (see, for example,
\cite{Swanson_95}). However,  spatio-temporal dynamics of
nutrients is not taken into account there.

In the present study a 1-D mathematical model for the infiltrative
tumour is developed. The model exploits the fact that tumour cell
division is possible only in the case of sufficient nutrient
concentration. Thus the account of the nutrient spatial
distribution effect on the tumour development is the central point
of the model considered. The model consists of reaction-diffusion
equations for the tumour cell density and nutrient concentration.
Dependence of the tumor front propagation rate on the model
parameters is obtained both analytically and numerically.
Properties of the propagating wave of the dividing tumour cells
described by the model are compared to the solution of the Fisher
equation.

\section{ The Model} \label{model}
We consider a tumour as a colony of living and dead cells
surrounded by normal tissue. Living cells divide with a constant
rate and can move in a random way. In the case of nutrients lack
tumour cells start to die.  We take into account in the model that
though a variety of nutrients are necessary for the tumour growth,
the oxygen shortage is mostly responsible for the cell death. We
consider a tumour growing in a normal tissue with rather poor
capillary system, so oxygen diffuse from the blood vessels located
sufficiently away from the tumour. Oxygen consumption by normal
cells which do not proliferate is neglected as dividing cells
consume nutrients much faster than non-proliferating ones. Normal
tissue surrounding the tumour is also supposed not to hinder
neither cancer cell motion nor proliferation. We restrict our
analysis to the case of a single spacial dimension. More formally
the model is based on the following physical and biological
assumptions

\begin{itemize}
\item A tumour grows as a colony of living and dead cells with
densities $a$ and $m$ correspondingly.
\item Living cells divide with a constant rate $B$ and in the case of low nutrient
concentration $s$ starve to death.
\item The death rate $P(s)$ depends on the nutrient concentration in a threshold
manner which is described below.
\item Only random motility (diffusion) with a constant coefficient $D_a$ of tumour cells is
considered in the model.
\item Oxygen is considered as a crucial nutrient for tumour growth. Its distribution
is governed by diffusion $D_s$ and consumption by the living
tumour cells only according to linear law $qa$.
\end{itemize}
Using these assumptions the governing equations can be written as
follows
\begin{equation}\label{PDE_3}
\begin{array}{l}
\displaystyle \frac{\partial a}{\partial t} = D_a \frac{\partial^2
a
}{\partial x^2}  - P(s)a + Ba,\\
\\
\displaystyle \frac{\partial m}{\partial t} =  P(s)a,\\
\\
\displaystyle \frac{\partial s}{\partial t} = D_s \frac{\partial^2
s }{\partial x^2}  - qa,
\end{array}
\end{equation}
where $a$, $m$ and $s$ are the proliferating cell density, the
dead cell density and the nutrient concentration correspondingly.
These equations are written already in a non-dimensional form. In
order to estimate the corresponding parameters we take
characteristic scales of time and length as $T_0 = 10^6$ s and
$L=5\cdot10^{-2}$ cm respectively. The cell density is scaled on
$a_{max}=10^7$ cells/cm$^3$ (maximal cell density) and the
concentration of oxygen in the tissue near blood vessels is
supposed to be $S_{max}=10^{-7}$ mol/cm$^3$. In dimensional values
the cell proliferation rate corresponds to the cell division
frequency of the order of one division per 100 hours. The
diffusion coefficients for oxygen and cells are equal to $D_s =
2.5\cdot10^{-5}$cm$^2/$s and $D_a = 2.5 \cdot10^{-9}$ cm$^2/$s
respectively. Thus we obtain the following non-dimensional
parameters of the problem
\begin{equation}\label{Par}
\begin{array}{llllllll}
D_a = 10,&& D_s = 10^4,&& B=0.1,&& P_{m} = 0.2,\\
S_{crit} = 0.3, && \epsilon = 0.01 && q = 1.0,
\end{array}
\end{equation}
which will be referred to as a standard parameter set.

The cell death rate is governed by $P(s)$ which is a step-like
function. For $s>S_{crit}$ it is almost equal to zero and for
$s<S_{crit}$ it is greater than $B$. We will take it in the form
\begin{equation}
P(s) = P_m \displaystyle \frac{1 - \tanh[(s -
S_{crit})\epsilon]}{2}.
\end{equation}
where $P_m$ is the maximal value of $P(s)$ and the parameter
$\epsilon$ defines the characteristic deviation of $s$ form
$S_{crit}$ at which the death rate changes form the values close
to zero to the values close to $P_m$.

In Eq. (\ref{PDE_3}) the second equation decouples from the rest
set. The dead cells density profile determines the position of
necrotic region inside the tumour. Therefore it does not affect the
tumour spreading, which is governed only by the dynamics of
proliferating cells and nutrient concentration and thus Eq.
(\ref{PDE_3}) is reduced to the  following equations
\begin{equation}\label{PDE}
\begin{array}{l}
\displaystyle \frac{\partial a}{\partial t} = D_a \frac{\partial^2
a
}{\partial x^2}  - P(s)a + Ba,\\
\\
\displaystyle \frac{\partial s}{\partial t} = D_s \frac{\partial^2
s }{\partial x^2}  - qa.
\end{array}
\end{equation}

At the avascular stage a tumour is supposed to have a spherically
symmetric shape. However if the radius of the tumour is much greater
than the characteristic scale,  for which the distribution of cell
density and nutrient concentration change significantly, then a
planar geometry can be considered. In view of our assumption that
the tumour grows in the tissue where oxygen predominantly diffuses
from the blood vessels located far away from the tumour, the set
(\ref{PDE}) can be solved in an infinite region. Thus we supplement
Eq. (\ref{PDE}) with the following boundary conditions on $a$ and
$s$
\begin{equation}\label{BC}
\begin{array}{llllll}
\left\{
\begin{array}{l}
a = 0,\\
s_x=0
\end{array}\right.&
\text{for} & x \rightarrow -\infty,& \left\{
\begin{array}{lll}
a=0,\\
s=S_{max}
\end{array}\right.&\text{for}&
x\rightarrow +\infty.
\end{array}
\end{equation}

Undoubtedly, any mathematical model of a biological system depends
crucially on the choice of parameter values. The parameters vary
drastically depending on tumour type, localization, progression
etc. The choice of parameters in the model is determined by our
objective which is rather a qualitative description of an
infiltrative tumour, but not a quantitative description of any
specific tumor. Therefore the parameter values are taken in the
experimentally observable range and they are not related to a
certain kind of cancer. The other important limitation on
parameters is that their values are chosen in such a way that the
tumour cell density remains substantially smaller than the maximal
value, which is typical for an infiltrative type of a tumour.

\section{Travelling wave solutions}\label{TWS}

We seek the solution to Eq. (\ref{PDE}) in the form of a wave
travelling with a constant shape and speed $c$ (i.e. an autowave).
Then the governing equations (\ref{PDE}) are reduced to a set of
ordinary differential equations by introducing the automodel
coordinate frame $\xi = x- ct$
\begin{equation}\label{ODE}
\begin{array}{l}
\displaystyle D_a \frac{\partial^2 a
}{\partial \xi^2} +  c\frac{\partial a}{\partial \xi}  - P(s)a + Ba = 0,\\
\\
\displaystyle D_s \frac{\partial^2 s }{\partial \xi^2} +
c\frac{\partial s}{\partial \xi}  - qa = 0
\end{array}
\end{equation}
with the following boundary conditions
\begin{equation}\label{BC_2}
\begin{array}{llllll}
\left\{
\begin{array}{l}
a = 0,\\
s = \sigma
\end{array}\right.&
\text{for} & \xi \rightarrow -\infty,& \left\{
\begin{array}{lll}
a=0,\\
s=1
\end{array}\right.&\text{for}&
\xi\rightarrow +\infty.
\end{array}
\end{equation}
Here the constant $\sigma$ corresponds to the limiting constant
value of the substrate concentration on $-\infty$. We refer to the
parameters $c$ and $\sigma$ as internal parameters contrary to the
control parameters, listed in (\ref{Par}). On the first step we
consider the asymptotic behavior of the solutions of Eq. (\ref{ODE})
on $\xi =\pm \infty$. In order to do this we linearize Eq.
(\ref{ODE}) near the values (\ref{BC_2}) which are stationary points
of (\ref{ODE}) and obtain two sets of ODEs with constant
coefficients
\begin{equation}\label{ODE-}
\begin{array}{l}
\displaystyle D_a \frac{\partial^2 a^-
}{\partial \xi^2} +  c\frac{\partial a^-}{\partial \xi}  (B- P(\sigma))a^- = 0,\\
\\
\displaystyle D_s\frac{\partial^2 s^- }{\partial \xi^2} +
c\frac{\partial s^-}{\partial \xi}  - qa^- = 0,
\end{array}
\end{equation}
and
\begin{equation}\label{ODE+}
\begin{array}{l}
\displaystyle D_a \frac{\partial^2 a^+
}{\partial \xi^2} +  c\frac{\partial a^+}{\partial \xi}  Ba^+ = 0,\\
\\
\displaystyle D_s \frac{\partial^2 s^+ }{\partial \xi^2} +
c\frac{\partial s^+}{\partial \xi}  - qa^+ = 0.
\end{array}
\end{equation}
According to linear differential calculus the solutions to these
problems are sought in the form $(a,~a_{\xi},~s,~s_{\xi})^T \sim
{\bf k}^{\pm} \exp(\mu^{\pm} \xi)$ which reduces the systems of
linear differential equations to eigenvalue problems for
coefficients $\mu^{\pm}$ as eigenvalues and constant vectors ${\bf
k}^{\pm}$ as eigenvectors. Eigenvalues $\mu^{\pm}$ indicate the
rate of exponential convergence (divergence) of the solutions to
the boundary values (\ref{BC_2}) as $\xi \rightarrow \pm \infty$
and superscripts `+' and `-' refer to the linearized problem on
$+\infty$ and $-\infty$ respectively.

For the case $\xi \rightarrow - \infty$ it appears that the
linearized set of ODEs (\ref{ODE-}) has a singe solution with the
rate of exponential convergence to the boundary conditions given by
$\mu^-_1 = (-c + \sqrt{c^2 + 4 (P(\sigma) -B) D_a})/2D_a$, two
solutions unbounded for $\xi \rightarrow -\infty$ with the rates of
exponential divergence $\mu^-_2 = (-c - \sqrt{c^2 + 4 (P(\sigma) -B)
D_a})/2D_a$, $\mu^-_3 =-c/D_s$, and a single solution with a zero
coefficient, $\mu^-_4 =0$, of exponential growth. It can be shown
that the stationary point $\{S1: a=0, s=\sigma\}$ of (\ref{ODE}) is
either a saddle-node if $\mu^-_1$ is real positive, a stable node if
$\mu^-_1$ is real negative, or a stable focus if $\mu^-_1$ is a
complex number.

In the same manner, for the case $\xi \rightarrow +\infty$ we derive
the set of ODEs (\ref{ODE+}) linearized near the boundary conditions
(\ref{BC_2}) which has four linearly independent solutions bounded
for $\xi \rightarrow +\infty$ and characterized by the rates of
exponential convergence $\mu_{1,2}^+ = (-c \pm \sqrt{c^2 - 4 B
D_a})/2D_a$, $\mu^+_3 = - c/D_s$, $\mu^+_4 = 0$. A simple
consideration shows that the stationary point $\{S2: a=0, s=1\}$ is
either a stable node or a stable focus. In the latter case $a$ can
become negative and therefore this solution is physically
unrealistic.

The eigenvectors associated with the linearized problems can be
written as ${\bf k}^{\pm}_{1,2} = (1,~\mu^{\pm}_{1,2},
q/\Lambda^{\pm}_{1,2},~\mu^{\pm}_{1,2} q/\Lambda^{\pm}_{1,2} )^T$,
${\bf k}^{\pm}_{3} = (0,~0,~1,\mu^\pm_{3})^T$ and ${\bf
k}^{\pm}_{4} = (0,~0,~1,~0)^T$. Here we have introduced a notation
$\Lambda^{\pm}_{1,2} =\sqrt{ D_s
\mu^{\pm2}_{1,2}+c\mu^\pm_{1,2}}$.

Taking into account that  the solution of Eq. (\ref{ODE},\ref{BC_2})
exist only if the coefficient $\mu^-_1$ is a real positive number
($S1$ is a saddle-node) and $\mu^+_{1,2,3}$ are real numbers ($S2$
is a stable node) we derive the following restrictions on the model
parameters
\begin{equation}\label{cond}
\begin{array}{lll}
P(\sigma) > B, && \displaystyle c \geq  2\sqrt{B D_a}.
\end{array}
\end{equation}
The last condition implies that an automodel solution of Eq.
(\ref{PDE}) can propagate only with the velocity higher or equal to
some minimal value $c_{min} = 2\sqrt{B D_a}$. This property of the
set (\ref{PDE}) solutions is similar to the one of the KPP
(Kolmogorov-Petrovskii-Piskunov) \cite{KPP37} equation and its
special case, Fisher equation \cite{Fisher37}, which also exhibit
autowaves.

The first inequality in (\ref{cond}) shows that since $P(s)$ is a
monotonic decreasing function there exist autowaves characterized
by the residual concentration of the substrate left behind the
travelling wave and this residual concentration is less or equal
to some critical value $\sigma_{max} = P^{-1}(B)$, i.e.
$\lim_{\xi\rightarrow - \infty} s(\xi) = \sigma \leq
\sigma_{max}$. The existence of both the parameter $\sigma$ and
the limiting condition makes our model different from the KPP
model.

It can be also shown that for the travelling wave solution of Eq.
(\ref{ODE}) the substrate concentration profile is a monotonic
function of the coordinate $\xi$. In order to do this we multiply
the second equation in (\ref{ODE}) by the integrating factor
$exp(c/D_s \xi)$ and integrate it over $\xi$ to obtain
\begin{equation}\label{monot}
D_s s(\xi)_{\xi} e^{c/D_s \xi} = q\int_{-\infty}^{\xi}
a(z)e^{c/D_s z}dz.
\end{equation}
The integral in the right hand side of Eq. (\ref{monot}) is always
positive therefore $s_{\xi}$ is positive and $s(\xi)$ is a monotonic
growing function.

\section{Numerical simulations}

We solve Eq. (\ref{ODE}) subject to boundary conditions (\ref{BC})
numerically by using shooting and relaxation methods. Here we skip
the description of these methods since they can be found in our
previous papers (see \cite{SIAM} and references therein). The
problem for numerical calculation is posed on a finite domain $\xi
\in [-L,L]$, where $L$ is taken to be sufficiently large, with a
uniform grid on it. The integration step and $L$ values are chosen
in such way that both twice decreasing the integration step and
twice increasing the integration interval results in a variation of
the calculated parameters (such as wave speed) in the ninth
significant digit only.

The stability analysis of autowave solutions travelling with
different velocities is studied numerically. Equations (\ref{PDE})
are solved in a coordinate frame moving with a minimal speed ($\xi
= x -c_{min}t$). The boundary conditions (\ref{BC}) are imposed at
the edge points of the space grid. For our numerical algorithm we
use the method splitting with respect to physical processes.
Initially we solve the set of ODEs which describes the cell birth
and death processes as well as nutrient consumption by using the
fourth order Runge-Kutta algorithm. As a next step, equations of
mass transfer for oxygen and cells are solved with the
Crank-Nicolson method of the second order approximation in space
and time.

\begin{figure}
\setlength{\epsfxsize}{70mm}\centerline{\epsfbox{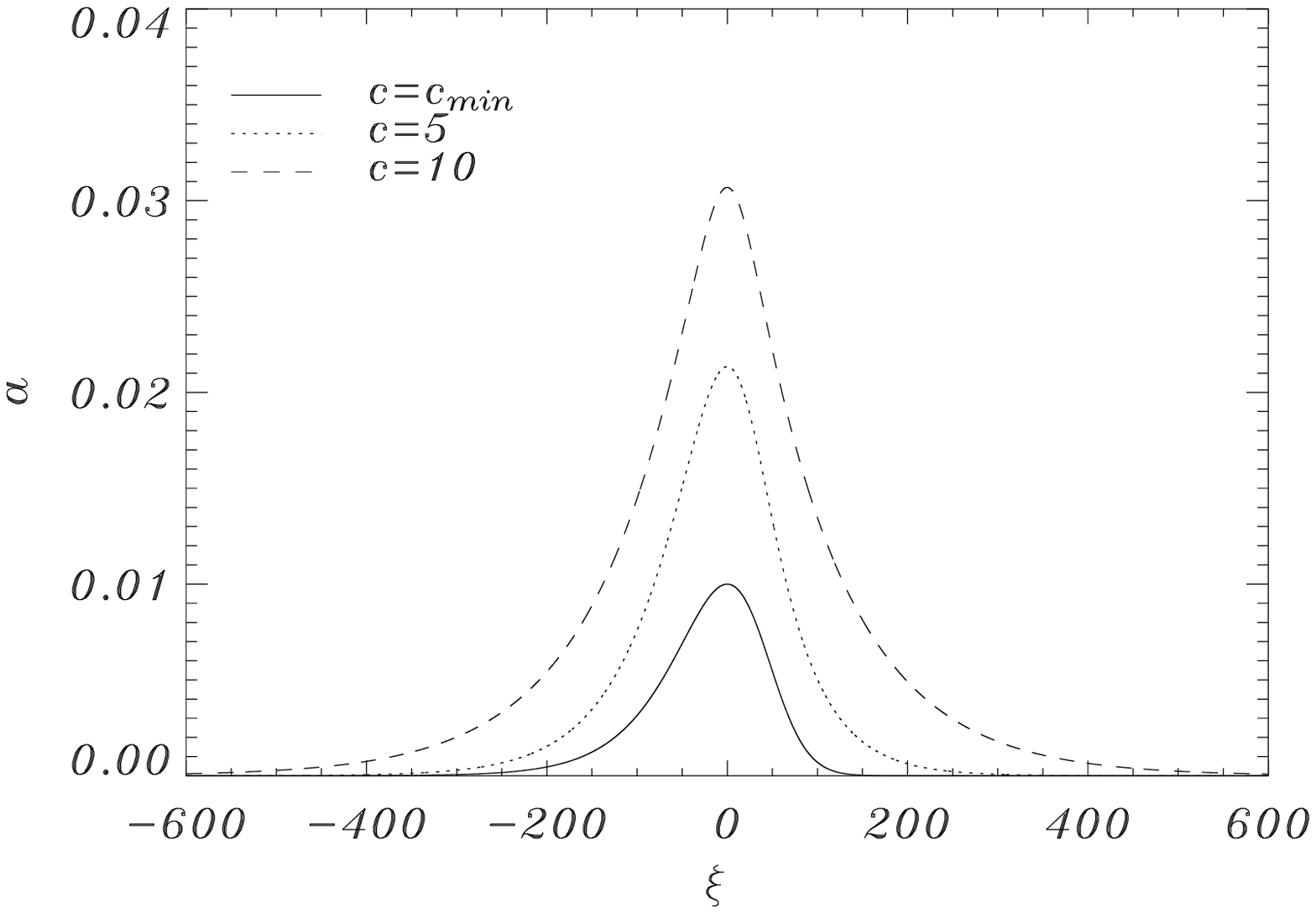}}
\setlength{\epsfxsize}{70mm} \centerline{\epsfbox{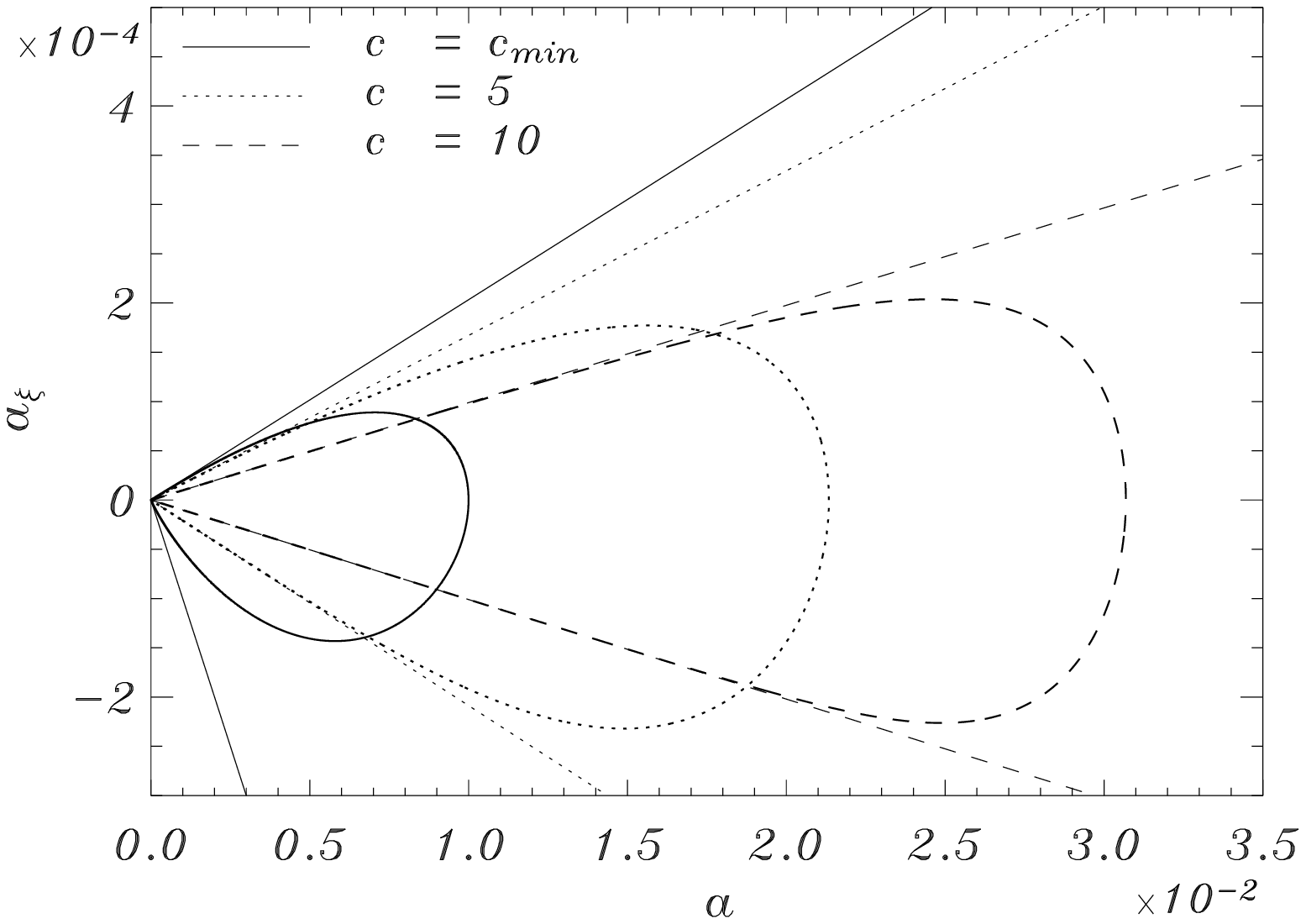}}
\caption{Cell density profile as a function of coordinate $\xi$ in
the moving frame (top figure) and projections of phase trajectories
onto the plane $(a,~a_{\xi})$ (bottom figure) for standard parameter
values and various values of the wave speed $c\geq c_{min}$.
}\label{prof}
\end{figure}

Typical solution profiles for the cell density, $a(\xi)$, and a
projection of phase trajectories into the plane $(a,~a_{\xi})$ are
shown in the top and bottom figures \ref{prof}  respectively for
various values of the wave speed $c\geq c_{min}$, where $c_{min} =
2\sqrt2$ in this case. In the bottom figure \ref{prof} we also
show the asymptotic behavior of the solutions of Eq.
(\ref{ODE}-\ref{BC_2}) for $\xi \rightarrow \pm \infty$ with
straight lines which represent the trajectories of problems
obtained form Eq. (\ref{ODE}) by linearizing it near the boundary
conditions (\ref{BC_2}). The cell density profile looks like a
sharp peak so that $a$ vanishes quickly form its maximum value in
a relatively thin region of the $\xi$ coordinate. As we increase
$c$ the $a(\xi)$ the profile becomes less localized and the
maximum value of the cell density $a(0)$ increases as well. It can
be shown, however, that $a(0)$ always remains less than one. The
substrate concentration profile $s(\xi)$ is a monotonic growing
function of coordinate $\xi$ (not shown here for the sake of
brevity), which is almost constant, $s(\xi) \sim \sigma$ for
$\xi<0$ and exhibits a slow exponential growth to its boundary
value ($s=1$ as $\xi \rightarrow \infty$) for $\xi>0$ with the
rate of growth $c/D_s$. The switching between these two regimes
occurs in the thin region near $\xi=0$ where $a$ reaches the
maximal value and coupling between the two equations in
(\ref{ODE}) is strong.

\begin{figure}
\setlength{\epsfxsize}{70mm}\centerline{\epsfbox{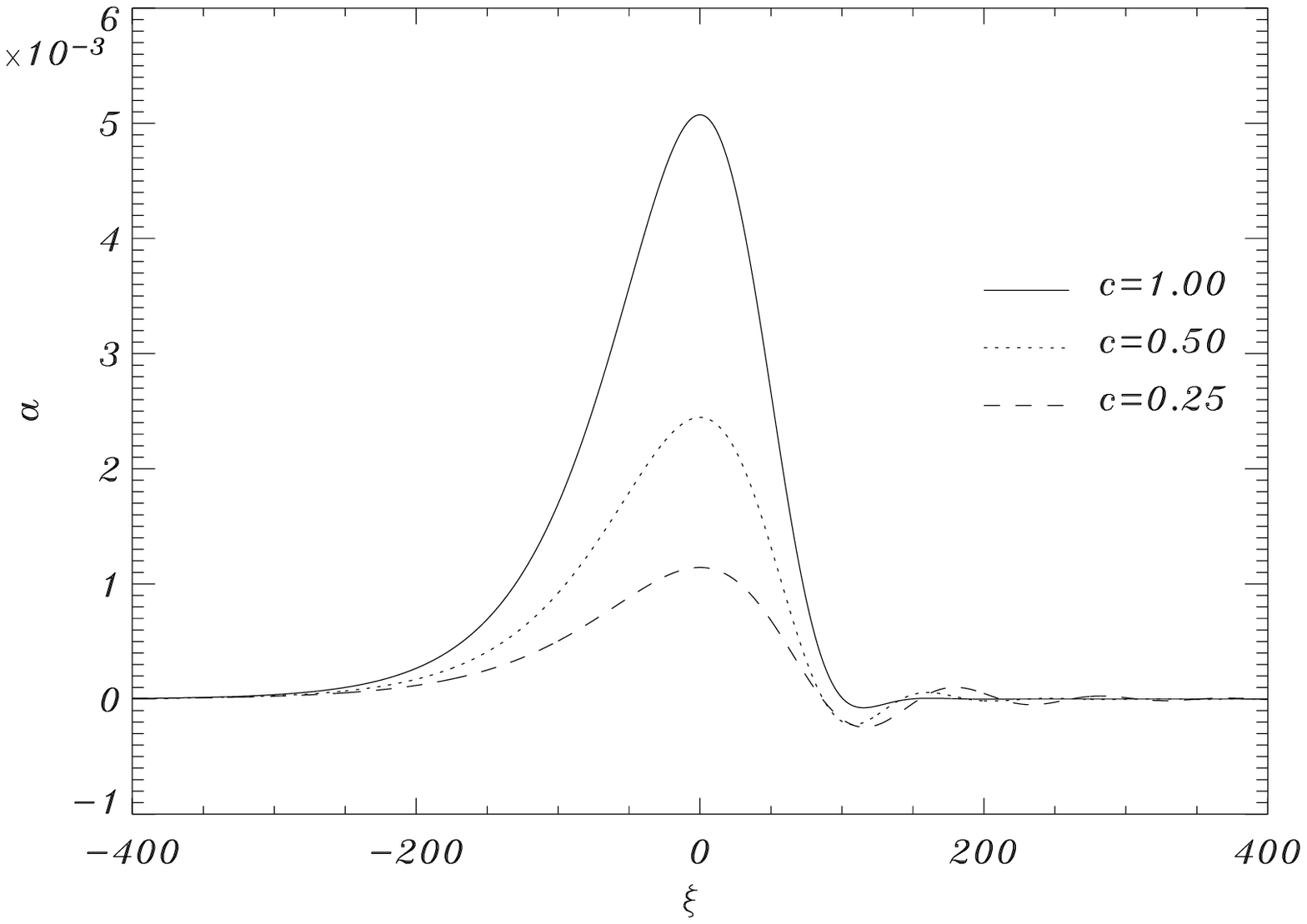}}
\setlength{\epsfxsize}{70mm} \centerline{\epsfbox{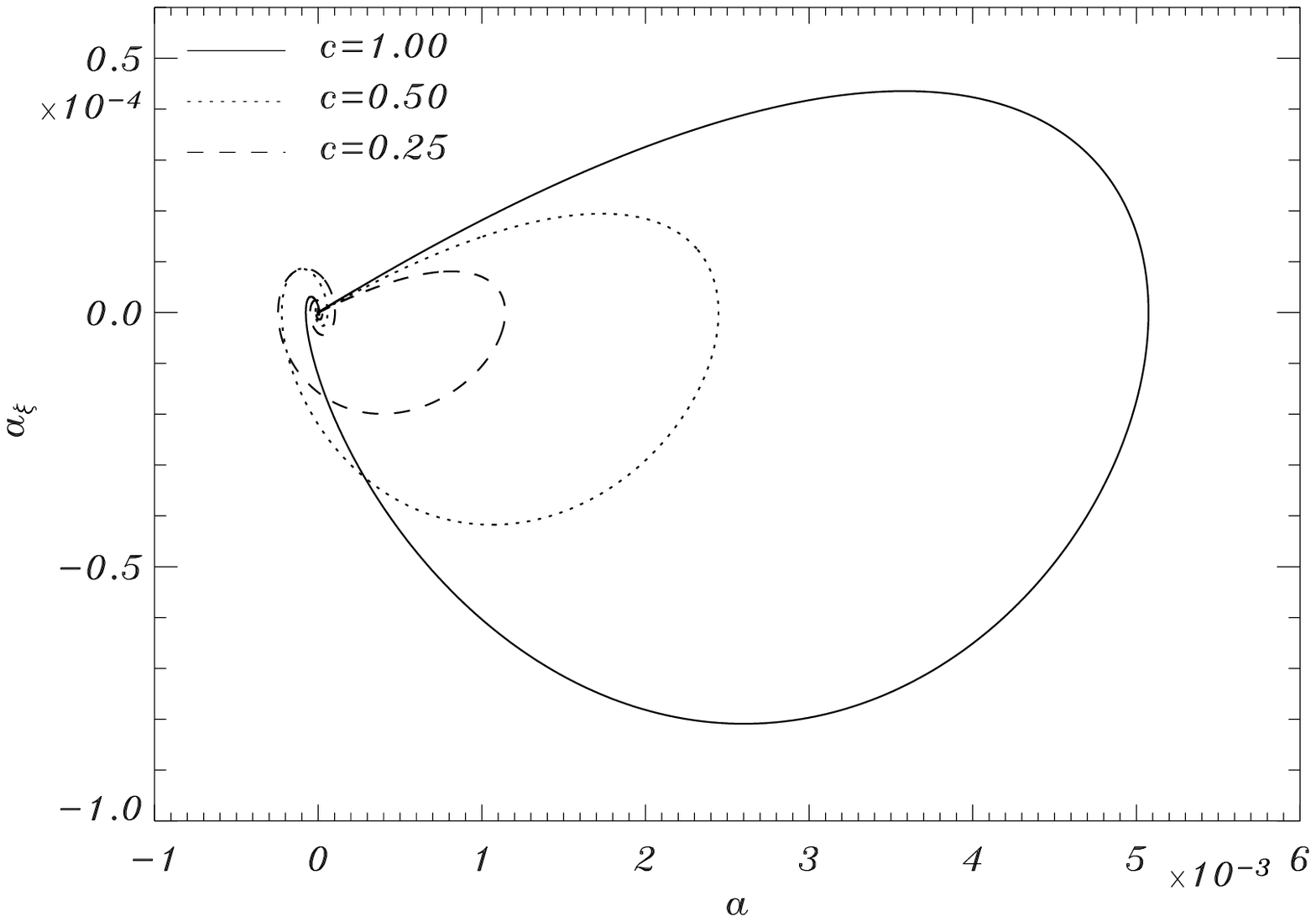}}
\caption{Cell density profile as a function of coordinate $\xi$ in
the moving frame (top figure) and projections of phase trajectories
onto a plane $(a,~a_{\xi})$ (bottom figure) for standard parameter
values and various values of speed $c < c_{min}$. }\label{prof_osc}
\end{figure}

In the top and bottom figures \ref{prof_osc} we plot a typical
solution profiles for the cell density $a(\xi)$, and a projection of
phase trajectories onto the plane $(a,~a_{\xi})$ respectively for
the several values of speed below the minimal value $c < c_{min} =
2$ as shown in figure. In figure \ref{prof_osc} it is clearly seen
that at certain $\xi$ values cell density becomes negative which is
not physically feasible. Therefore we consider the solutions for
which $a(\xi)\geq 0$, travelling with the speed higher than or equal
to a minimal value $c_{min}$.

\begin{figure}
\setlength{\epsfxsize}{70mm} \centerline{\epsfbox{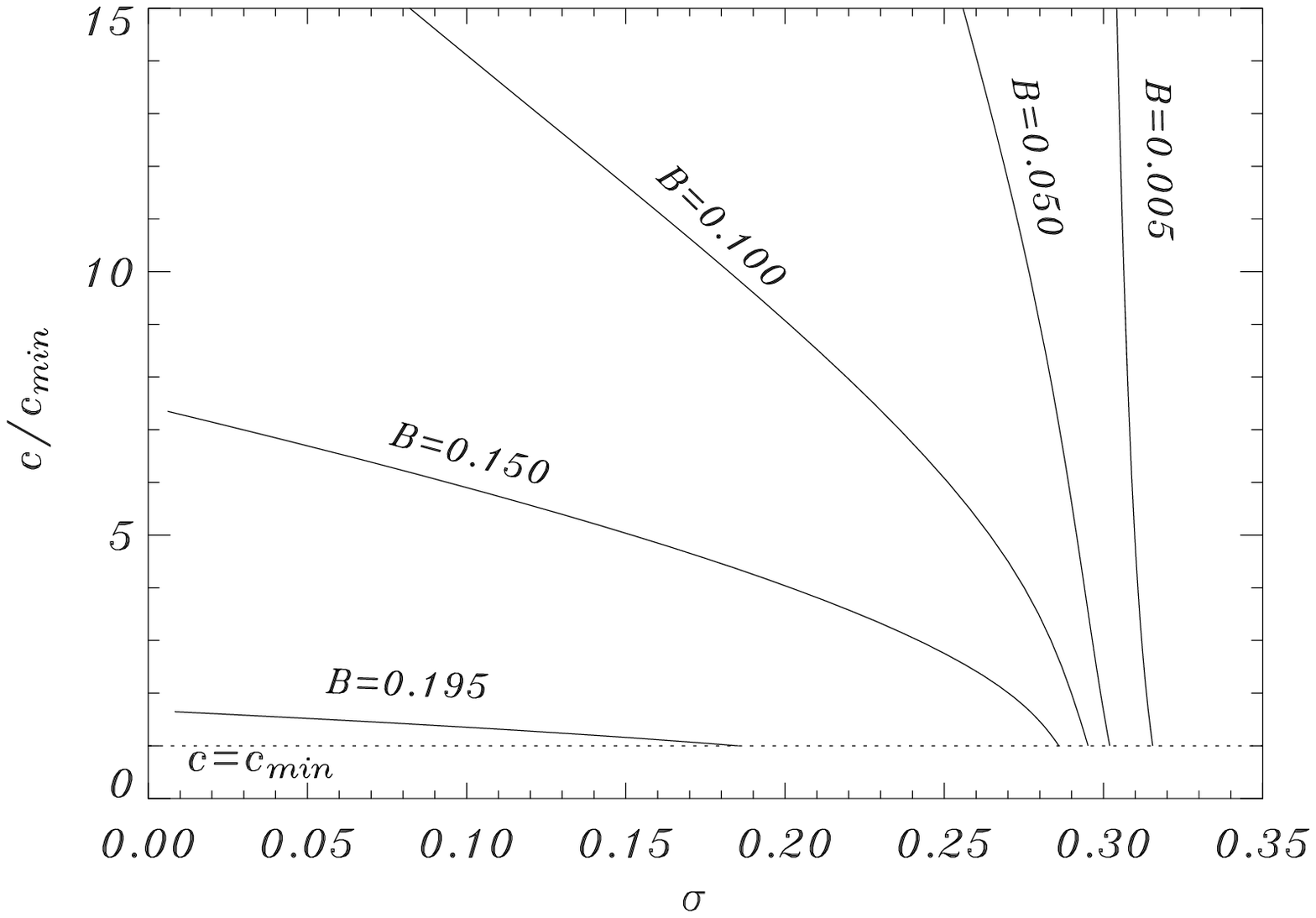}}
\setlength{\epsfxsize}{70mm} \centerline{\epsfbox{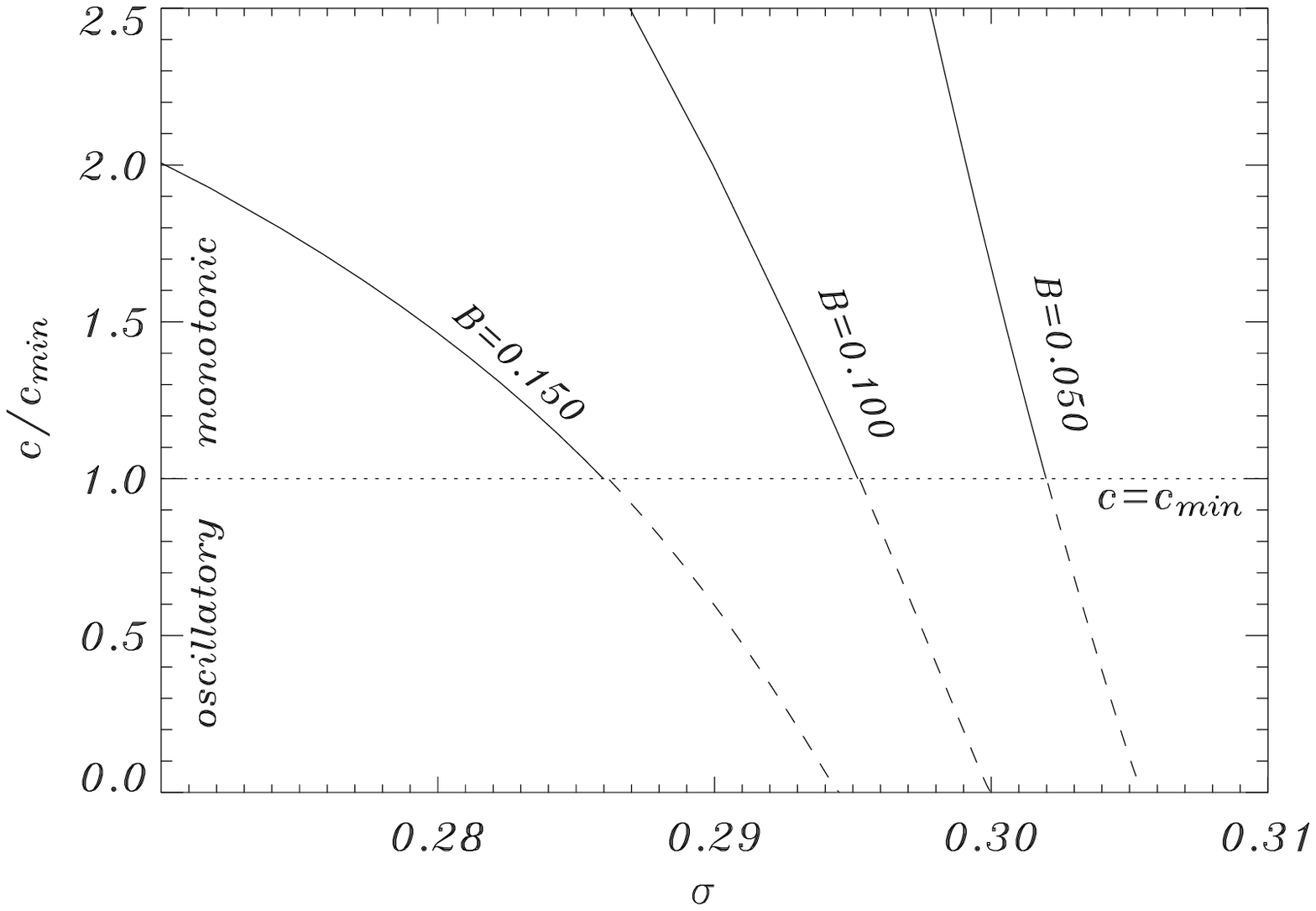}}
\caption{Dependence of $c/c_{min}$ on $\sigma$ for various values
of $B$.}\label{c(sigma)}
\end{figure}

As a next step we investigate the dependence of the speed of the
autowave solution as a function of parameters of the problem. In the
top figure \ref{c(sigma)} the dependence of $c$ on $\sigma$ is
plotted for the physically feasible solutions travelling with the
speed higher than $c_{min}$ (we refer to them as monotonic
solutions) for standard parameter values and several values of $B$
as shown in Fig. \ref{c(sigma)}. For fixed control parameter values
there exists a family of solutions $c(\sigma)$ travelling with
different speeds. For each value of $B$ the speed is scaled on the
minimal value $c_{min} = 2\sqrt{B D_a}$, which is also shown with a
dotted line marked "$c=c_{min}$". All monotonic solution branches
cease to exist when $c/c_{min}$ ratio drops down to $1$. Below this
value a family of the oscillatory solution (which are not physically
feasible) branches emerges. These solutions are plotted in the
bottom figure \ref{c(sigma)} with dashed lines. It is worthwhile
noting that at $c/c_{min} = 1$ only the second condition in
(\ref{cond}) is violated, whereas the first condition is approached
along the oscillatory branches when speed of the wave drops down to
zero and $\sigma$ reaches maximum which is shown in the bottom
figure \ref{c(sigma)}.

\begin{figure}
\setlength{\epsfxsize}{70mm} \centerline{\epsfbox{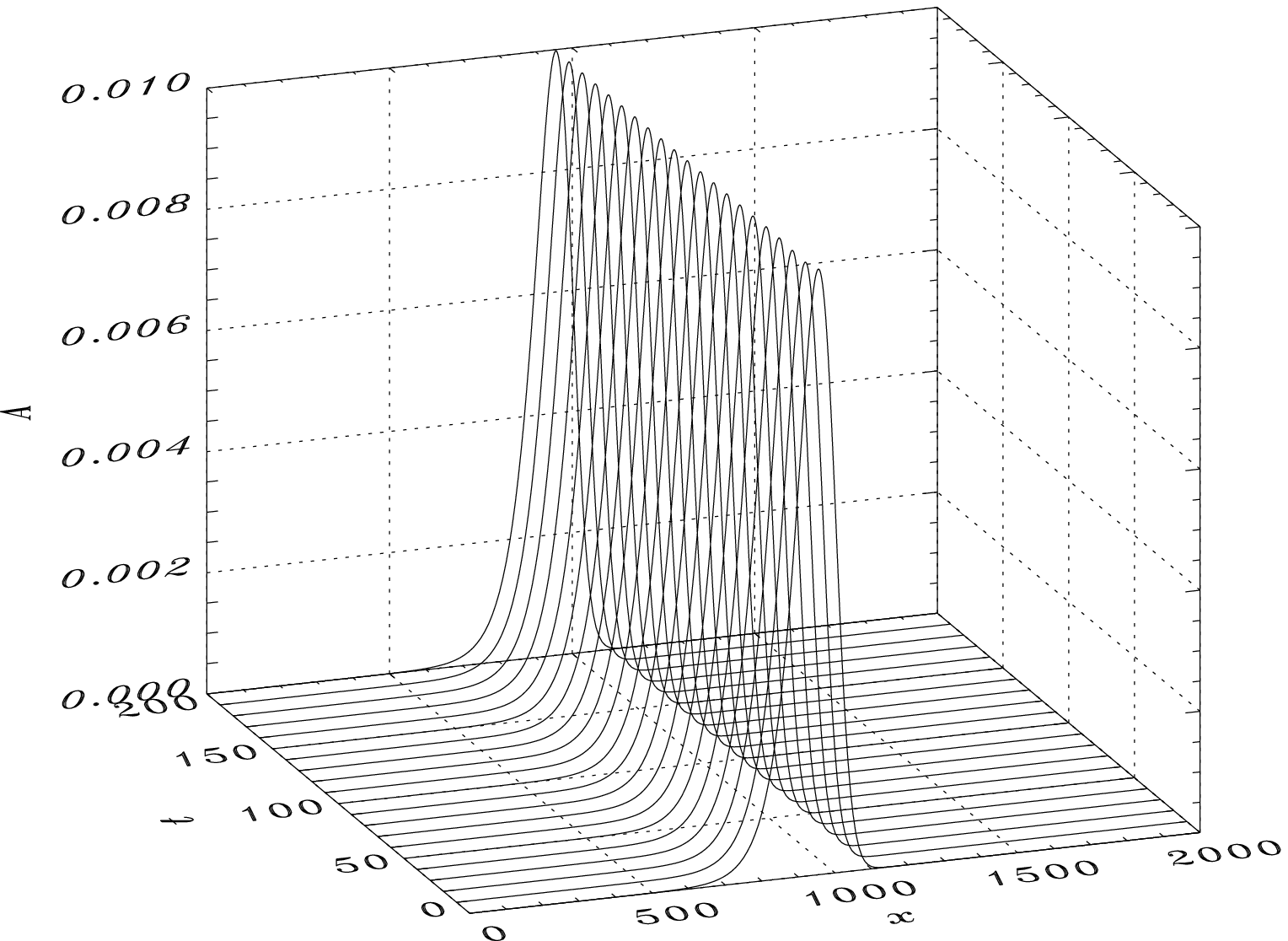}}
\setlength{\epsfxsize}{70mm} \centerline{\epsfbox{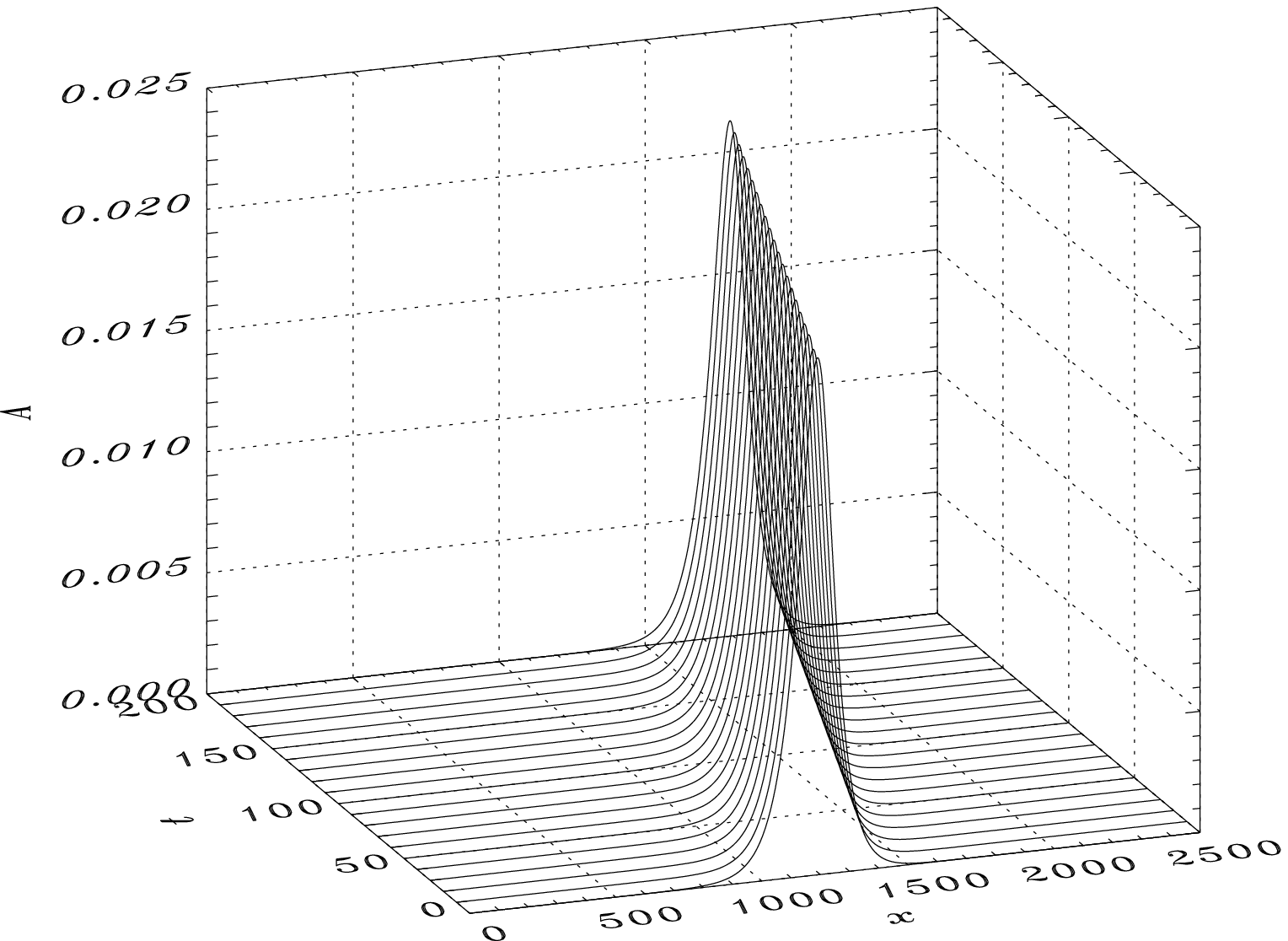}}
\caption{Dependence of cell density, $a$, on coordinate, $\xi$, in
the frame travelling with the minimal speed, and on time $t$ for the
standard parameter values. Initial profiles correspond to $c=
c_{min}$ (top figure) è $c= 5$ (bottom figure).}\label{A_min}
\end{figure}

We use the solutions obtained by solving Eq. (\ref{ODE}) as the
initial conditions for numerical integration of Eq. (\ref{PDE})
subject to boundary conditions (\ref{BC}) in order to investigate
the stability of these autowaves. In Fig. \ref{A_min} we show a
spatial-temporal evolution of the initial profiles obtained by
numerical integration of Eq. (\ref{ODE}) with the speed equal to
the minimal value, $c=c_{min}$ (top figure), and with the speed
higher than minimum, $c=5$ (bottom figure). The standard parameter
values are used to perform this calculation. We use the travelling
coordinate frame, $\xi = x - c_{min} t$, so that the solution
propagating with the minimal speed is presented in the top figure
\ref{A_min} as a standing wave, whereas the solution corresponding
to $c=5$ is shown in the bottom figure \ref{A_min} as a wave
travelling with the speed $c = 5-c_{min}$. The set (\ref{PDE}) is
integrated over the time periods of the order of $t=400B^{-1}$,
where $B^{-1}$ is the characteristic time scale for the
instability growth in Eq. (\ref{PDE}). Results of numerical
simulation of Eq. (\ref{PDE}) are shown in Fig. \ref{A_min} for
two values of the wave speed and demonstrate that autowaves can
propagate stably for long intervals of time without changing their
speed and form.

\section{Limit $\epsilon = 0$}

\begin{figure}
\setlength{\epsfxsize}{70mm}
\centerline{\epsfbox{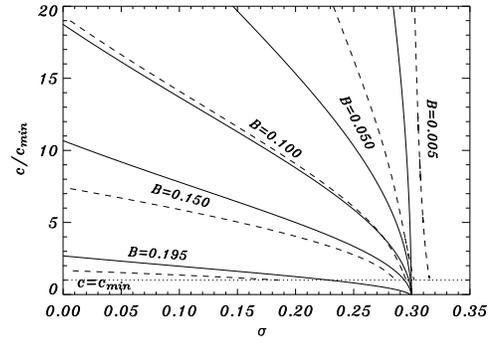}} \caption{Dependence of
$c/c_{min}$ on $\sigma$ for various values of $B$. The solid line
represents the results obtained from Eq.~(\ref{sigma(c)}) and the
dashed line shows the numerical data.}\label{c_step(sigma)}
\end{figure}

In the limit $\epsilon \rightarrow 0$ the governing equations
(\ref{PDE}) allow the following simplifications: (i) the cell
death rate function can be replaced by a Heaviside step function
$P_m H(s-S_{cr})$, (ii) due to translational invariance of
Eq.~(\ref{PDE}) we may require the travelling solution to satisfy
the condition $s(\xi)|_{\xi=0} = S_{crit}$. This implies that
$P(s(\xi)) = 0$ for $\xi<0$ and  $P(s(\xi)) = P_m$ for $\xi>0$. In
this case Eq.~(\ref{ODE}) can be considered in two semi-infinite
intervals $\xi \in(-\infty, 0)$ and $\xi \in (0, \infty)$
separately. This yields a set of ODEs similar to Eq.~(\ref{ODE-})
for $\xi<0$, where $P(\sigma)$ is replaced by $P_m$, and a set
(\ref{ODE+}) for $\xi>0$. The unknown functions $a^{\pm}$ and
$s^{\pm}$ satisfy the boundary conditions (\ref{BC_2}) for
$\xi\rightarrow \pm \infty$ respectively and can be written
explicitly as
\begin{equation}\label{sol}
{\bf u}^{\pm}(\xi) = \displaystyle \sum_{i=1}^4 \alpha^{\pm}_i
{\bf k}^{\pm}_i e^{\mu^{\pm}_i \xi},
\end{equation}
where ${\bf u}^{\pm}\equiv (a^{\pm},~
a_{\xi}^{\pm},~s^{\pm},~s_{\xi}^{\pm})^T$; ${\bf k}^{\pm}_i$ and
$\mu^{\pm}_i$ are taken from the section~\ref{TWS} (where
$P(\sigma)$ is replaced by $P_m$) and $\alpha^{\pm}_i$ are
constant coefficients which are to be found. From boundary
conditions (\ref{BC_2}) it follows that $\alpha^-_2 = 0$,
$\alpha^-_3 = 0$, $\alpha^-_4 = \sigma$ and $\alpha^+_4 = 1$. The
other four constants can be found from continuity conditions
across $\xi=0$ for $a(\xi)$: $a^-(0) = a^+(0)$, $s(\xi)$:  $s^-(0)
= s^+(0)$, and $s_{\xi}(\xi)=S_{crit}$:  $s_{\xi}^-(0) =
s_{\xi}^+(0)$. This yields four equations
\begin{equation}
\begin{array}{l}\label{s_syst_1}
\alpha^-_1 = \alpha^+_1 +  \alpha^+_2,\\
\\
\sigma + \displaystyle \frac{q}{\Lambda^-_1} \alpha^-_1 = 1 + \frac{q\alpha^+_1}{\Lambda^+_1}  + \frac{q\alpha^+_2}{\Lambda^+_2} + \alpha^+_3= S_{crit},\\
\\
\displaystyle \frac{q}{\Lambda^-_1} \alpha^-_1 \mu^-_1 =
\frac{q\alpha^+_1}{\Lambda^+_1} \mu^+_1 +
\frac{q\alpha^+_2}{\Lambda^+_2} \mu^+_1 + \alpha^+_3 \mu^+_3.\\
\end{array}
\end{equation}
Our aim is to find the dependence of the speed of the autowave
solution on parameters of the problem i.e. both the control
parameters (\ref{Par}) and the internal parameter $\sigma$. The
four unknown coefficients $\alpha^{\pm}_i$ together with $\sigma$
make five unknowns. In order to find $c$ we need one more equation
which can be obtained by integration of the second equation from
Eq.~(\ref{ODE}) over $\xi$ with infinite limits and taking into
account (\ref{BC_2}) and (\ref{sol}).
\begin{equation}\label{s_syst_2}
\begin{array}{l}
\displaystyle \int^{\infty}_{-\infty} a(\xi) d\xi= \frac{\alpha^-_1}{\mu^-_1} + \frac{\alpha^+_1}{\mu^+_1}  + \frac{\alpha^+_2}{\mu^+_2} = \frac{1-\sigma}{q},\\
\end{array}
\end{equation}
The set of five equations (\ref{s_syst_1}-\ref{s_syst_2}) is
linear with respect to $\alpha^-_1$, $\alpha^+_1$,  $\alpha^+_2$,
$\alpha^+_3$, and $\sigma$. It yields
\begin{equation}\label{sigma(c)}
\sigma = \displaystyle \frac{2S_{crit}c + 2S_{crit}D_s \mu_1^-
-c-1}{2D_s \mu_1^- + c-1 },
\end{equation}
where $\mu_1^-$ is taken for $\sigma=0$. The dependencies given by
Eq.~(\ref{sigma(c)}) are presented in Fig.~\ref{c_step(sigma)}
with solid lines. The speed of the front is plotted vs $\sigma$
for various values of $B$. For each $B$ the graph $c(\sigma)$ is
scaled onto its critical value $c_{min} = 2\sqrt{D_aB}$, which is
shown in the figure with the dotted line marked as $c/c_{min}=1$.
The solutions travelling with the speed $c\geq c_{min}$ are
described by Eq.~(\ref{sol}). For $\xi>0$ all coefficients
$\mu_{1-3}^+$ are real and positive, therefore $a$ and $s$
monotonically approach the limiting values (\ref{BC_2}) as $\xi$
tends to infinity. The solutions travelling with $c<c_{min}$ have
complex coefficients $\mu_{1-3}^+$ with negative real parts,
therefore for $\xi>0$ they exhibit oscillatory behavior and for
certain values of $\xi>0$ the cell density is negative contrary to
reality . The dependence of $c/c_{min}$ on $\sigma$ is also shown
in the same graph by the dashed line for the data obtained by
numerically solving of Eq.~(\ref{ODE}) . The results obtained
numerically and analytically are in good qualitative agreement.
Quantitatively the influence of the finite length of the $P(s)$
intermediate region (where it changes from $P_m$ to zero) is
stronger for small values of $B$, as well as $B$ close to $P_m$,
whereas it is moderate for $B$ close to $P_m/2$.

\section{Approaching the asymptotic behaviour}

\begin{figure}
\setlength{\epsfxsize}{90mm} \centerline{\epsfbox{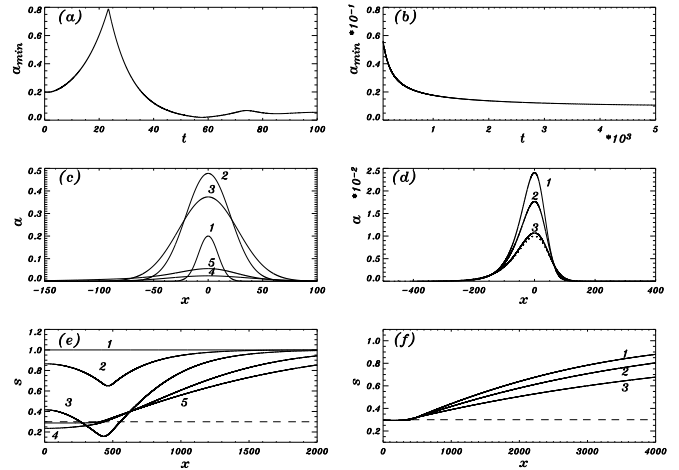}}
\caption{Dependence of maximal cell density, $a_{max}$, on time,
$t$, (figures (a) and (b)), the cell density profiles, $a(x)$,
(figures (c) and (d)) and the substrate concentration profiles,
$s(x)$, (figures (e) and (f)) sampled at various moments of time.
The cell density and nutrient concentration profiles are plotted in
the coordinate frame travelling with $c_{min}.$ The dependencies
$a(x)$ are centered at the origin so that $a(0)$ is equal to a
maximal value of $a$ over all $x$ values for fixed $t$.}\label{sum}
\end{figure}

It is known that for the Fisher equation the initially localized
profile evolves with time in such a way that it approaches the
automodel solution travelling with the minimal speed. Since the
model considered in this paper has much in common with the Fisher
equation, one may expect that it exhibit similar type of
behaviour. Here we present the results of the investigation of the
asymptotic behaviour of the solution of Eq.~(\ref{PDE}). The
uniform distribution of the nutrient concentration, $s=1$, and
gaussian distribution of the living cell density with the
characteristic width $\Delta = 20$ and amplitude $0.1$ has been
taken for the initial conditions (Fig.~\ref{sum}). It has been
shown that the process of the evolution of the initial profile can
be split into two stages: initial and relaxation ones.

The initial stage is shown in Fig.~\ref{sum} (a), (c) and (e),
where the maximal value of the cell density is plotted vs. $t$ in
figure (a) for the time interval $[0,~100]$, the cell density and
nutrient concentration profiles are shown for fixed moments of
time: $t_1=0$ (curves 1), $t_2=17$ (curves 2), $t_3=30$ (curves
3), $t_4=55$ (curves 4), and $t_5=100$ (curves 5) in figures (c)
and (e) respectively. Initially there is an excess of oxygen so
that the increase of cell density due to division exceeds density
decrease via both diffusion and cell death, which is negligible on
this  stage. Thus exponential growth of $a_{max}$ is observed.
Fast growth of the cell density is obviously accompanied with
rapid consumption of nutrients in the regions where $a$ is high.
As a result $s$ drops down below the critical value $S_{crit}$ and
the death rate $P(s)$ hops from zero to its maximum value $P_m$ in
the regions where $(S_{crit}-s)\gg \epsilon$. This case is
represented by the curve 3 in Fig.~\ref{sum}(e), where the
critical value of $s=S_{crit}$ is plotted by the dashed line. At
this stage cells cannot reproduce themselves effectively and cell
density decreases due to high death rate which is becoming the
dominating factor in the evolution of the $a$ profile in the
region close to $a_{max}$. It changes the trend for $a_{max}$ from
the exponential growth to practically exponential decay with the
index $P_m-B$ resulting in the appearance of sharp peak in the
$a_{max}(t)$ graph for $t\approx 23$. As a result the nutrient
consumption decreases, the nutrient concentration profile flattens
due to diffusion from the regions, where it has not been consumed
yet, and becomes monotonic. The cell density profile decays as is
shown by curves 4 and 5 in Fig.~\ref{sum}(e). For $t\in [50,100]$
there are oscillations in the $a_{max}$ value are observed due to
redistribution of oxygen and cells to and from the region where
$a\sim a_{max}$. These oscillations vanish for $t>100$, where the
regime of relaxation to the automodel solution starts. The
relaxation regime is characterized by a monotonic decay of
$a_{max}$ to the maximal value of cell density for the automodel
solution travelling with the minimal speed. This is illustrated in
Fig.~\ref{sum} (b), (d) and (f), where the dependence of $a_{max}$
on $t$ (in figure (b)) and the solution profiles $a(x,t)$ and
$s(x,t)$ (figures (d) and  (f) correspondingly) are plotted for
several moments of time $t_1 = 100$, $t_2= 1000$ and $t_3 = 5000$.
In figure (d) we also plot by the dotted line the automodel
solution profile travelling with the minimal speed. The dashed
line in figure (f) corresponds to $s=S_{crit}$. Just as in case of
the Fisher-KPP equation \cite{Saar} the relaxation of the cell
density profile to automodel solution travelling with the minimal
speed follows the power-mode law. However in our case the
characteristic time of relaxation is proportional to $D_s$, what
indicates that the dynamics of nutrient is important for adequate
description of the tumour evolution.

\begin{figure}
\setlength{\epsfxsize}{90mm}\centerline{\epsfbox{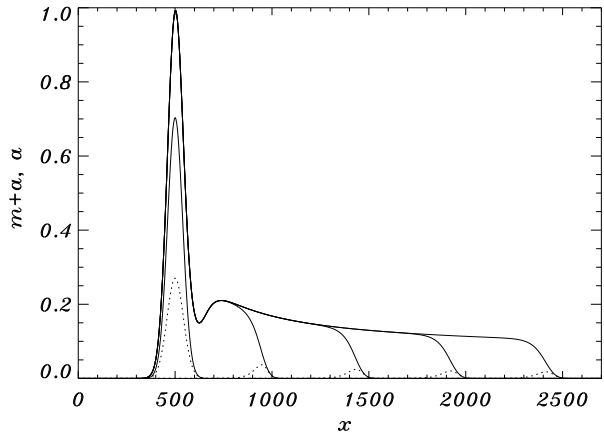}}
\caption{The overall cell density concentration profiles $a+m$
(solid line) and the living cell density profiles $a$ (dotted line)
vs coordinate $x$ sampled at various moments of time: $t = 50$,
$250$, $500$, $725$ and $1000$.}\label{A-B}
\end{figure}

To complete the description of the tumour growth process, the full
set of the governing equations (\ref{PDE_3}) was solved numerically.
In Fig.~\ref{A-B} the overall cell density distribution $a + m$,
which is comprised of both living $a(x,t)$ and dead $m (x,t)$ cells
is shown. In this figure the active cell density $a$ is plotted by
the dotted line and the overall cell density $a+m$ -- by the solid
line. The profiles are sampled at fixed moments of time $t = 50$,
$250$, $500$, $750$, and $1000$. The initial conditions are similar
to those, taken for calculations shown in Fig.~\ref{sum} i.e. the
substrate concentration is uniform in space and $s(x,0)=1$, the
living cell density $a$ is taken in a form of a gaussian
distribution of the width $\Delta = 100$ and amplitude $0.01$ and
the density of necrotic cells equals zero, $m(x,0) = 0$. The
dynamics of the gaussian profile $a(x,t)$ and the corresponding
dynamics of the substrate concentration are described earlier in
this section. Therefore here we discuss how this dynamics correlates
to the evolution of the overall tumour cell distribution and density
of necrotic cells. At the initial stage while living cell density
grows almost exponentially with time the density of dead cells
remains negligible. As tumour grows, the nutrient concentration
drops below the critical value $S_{Crit}$ in the region where the
majority of living cells are located. This results in the change of
trend from exponential growth of living cell density to its almost
exponential decay due to cell death. This process is obviously
accompanied by fast accumulation of dead cells. The density $m$
starts to grow rapidly so that a sharp peak in the $m(x)$
distribution appears. In the course of this process most of living
cells in the primary tumour site die, forming the necrotic region
shown by a sharp peak in Fig.~\ref{A-B}.

A small fraction of living cells, that survived at this stage,
spreads into surrounding tissue moving towards the source of
nutrient. This results in the formation of a travelling wave of
active cells on the tumour border. In the interior region a plateau
of the dead cell $m(x)$ density arises as a result of the living
cell death due to the shortage of oxygen. As the profile of the
living cells density approaches the automodel solution, the dead
cell density behind the front tend to the stationary value
$m_{stat}$. In the case of $\epsilon \rightarrow 0$ this value is
given by:

\begin{equation}\label{m_st}
\begin{array}{l}
 m_{stat} = \frac{P_m}{c_{min}}\displaystyle \int^{\infty}_{-\infty} a(\xi) d\xi,\\
\end{array}
\end{equation}

where $a(\xi)$ is the living cells density profile, corresponding to
the automodel solution of Eg.~(\ref{PDE}) traveling with the minimal
speed $c_{min}$.

The tumour cell density spatial distribution for $t=500$, $750$ and
$1000$ is shown in Fig.~\ref{A-B}. The total malignant cell density
$a+m$ is close to the maximal possible value, equal to unity, near
the tumour origin $x=500$. It can be interpreted as the formation of
a solid necrotic core in the primary tumour site. It is clearly seen
that tumour grows only towards the nutrient source and outside the
primary site the overall malignant cell density is substantially
less than unity. Thus tumour rather infiltrates neighbour normal
tissue with a constant speed but not grows like a solid object.

\section{Discussion}

A mathematical model for the infiltrative tumour growth which
includes distribution of living cells and oxygen in tissue is
developed. The model adequately describes a constant rate of the
tumour linear size growth at the initial stage of neoplasm
development and the formation of necrotic region in the tumour
interior, observed in experiments \cite{Freyer_86}.

In this model the existence of a family of autowave solutions
travelling with different velocities is demonstrated. Their
properties are investigated both numerically and analytically. It is
shown that for fixed parameter values the autowave solutions can
propagate with speeds higher or equal to some critical minimal value
$c_{min}$. In this sense our model is similar to the KPP-Fisher
model. However it has an essential new feature, compared to the
latter, due to the presence of the second equation. Namely there is
a close connection between the automodel wave velocity $c$ and the
residual nutrient concentration $\sigma$ behind the front.

The evolution of the initially localized profile is investigated. It
is shown that initially localized distribution asymptotically
approaches the automodel solution traveling with the minimal speed
according to power-mode law. This again resembles the behavior of
the autowave solutions of the Fisher equation. However in contrast
to Fisher model in our case the rate of the convergence depends
strongly on the parameters of the equation describing the evolution
of nutrient. From biophysical point of view the convergence of the
initially localized profile to an autowave solution implies that if
there is an initial center of tumor growth with nonzero density of
active cells, it will eventually develop into a full scale neoplasm
with a definite rate of linear growth of its size, accompanied by
formation of necrotic region inside of it. Our investigation shows
that this process can be divided into three stages. The first stage
is characterized by a fast growth of the tumour in the region with
non-zero initial tumour cell density. This stage lingers until
nutrient concentration drops below the critical value. The first,
fast growth stage is followed by second, intermediate, stage with a
rapid death of living cells in the tumour original position and
appearance of necrotic region there. The second stage lasts until
the nutrient consumption by the tumour gets counterbalanced by its
diffusion from the external source. Finally, the travelling wave
pattern which evolves to the automodel solution is formed and the
active tumour cells start to spread with constant velocity towards
the source of nutrients. Properties of the automodel solution are
basically determined by the equation for the malignant living cell
density and have much in common with automodel solution in the
KPP-Fisher equation. It should be noted that in real biological
systems with finite size autowave regime can be unattainable due to
the lack of time and/or  due to influence of the boundary.

In the present study we focused our attention on tumours of
infiltrative type, in which malignant cell density is substantially
smaller than the maximum cell density in tissue. In this case it
appeared possible to consider a relatively simple model, which takes
into account only individual cell motility and disregards convective
fluxes arising due to cell division, essential for solid tumours.
Actually, model simulations demonstrate that with the exception of
the primary site of tumour growth the total malignant cell density
is substantially smaller than the maximum cell density in tissue.
What is more important, the density of dividing cells, which gives
rise to convective fluxes in solid tumours, constitutes only several
percent of the maximal tissue density. Obviously,  with the other
set of model parameters this condition may not be fulfilled. Since
the tumour cell distribution tends to the automodel solution we were
able first to investigate its properties and to determine whether
the tumour belongs to the infiltrative type and thus the simplified
model can be used, otherwise convective fluxes should be taken into
consideration.

%

\begin{acknowledgments}
The authors acknowledge the financial support from RFBR: grants
05-01-00339 and 05-02-16518 and Russian Academy of Sciences: program
"Problems in Radiophysics".
\end{acknowledgments}

\bibliography{paper7}
\end{document}